\documentstyle[prl,aps,multicol,epsf]{revtex}
\renewcommand{\narrowtext}{\begin{multicols}{2}
\global\columnwidth20.5pc} 
\renewcommand{\widetext}{\end{multicols}
\global\columnwidth42.5pc} \multicolsep = 8pt plus 4pt minus 3pt

\newcommand{\be}{\begin{equation}}
\newcommand{\ee}{\end{equation}}

\begin{document}
\draft
\title{Consequences of a possible adiabatic transition between 
$\nu=1/3$ and $\nu=1$ quantum Hall states in a narrow wire.}

\author{Dmitri B. Chklovskii and Bertrand I. Halperin}
\address{Lyman Laboratory of Physics, Harvard University, Cambridge, MA
02138.}

\date{\today} 
\maketitle

\begin{abstract}
We consider the possibility of creating an adiabatic transition
through a narrow neck, or point contact, between two different
quantized Hall states that have the same number of edge modes, such as
$\nu=1$ and $\nu=1/3$.  We apply both the composite fermion and the
Luttinger liquid formalism to analyze the transition. We suggest that
using such adiabatic junctions one could build a DC step-up transformer,
where the output voltage is higher than the input.  Difficulties
standing in the way of an experimental implementation of the adiabatic
junction are addressed.
\end{abstract}


\narrowtext

It has long been understood that quantized Hall states with different
Hall conductances cannot be connected adiabatically in the interior of
a macroscopic two-dimensional electron system.  For a pure system,
where the quantized Hall states have energy gaps, the boundary between
two quantized states must be characterized by a vanishing energy gap,
with associated low energy excitations.  In a disordered system there
are generally localized low energy excitations in the interior of a
quantized Hall region, which then become extended at the boundary
between two quantized regions.  The possible transitions between
different quantized Hall states have been elucidated (in the case of a
fully spin-polarized system) by the introduction of a "global phase
diagram" based on a unitary transformation which introduces a
Chern-Simons gauge field and which, at the mean field level, maps
fractional quantized Hall states onto integer ones. \cite{KLZ,HLR}

In this Letter, we suggest that in a {\it narrow quantum wire} there
can be an adiabatic transition between two different quantized Hall
states, under certain conditions.  The most important example, to
which we restrict ourselves here, is the case of a transition between
states with $\nu = 1$ and $\nu = 1/3$.  It should be noted that for
both these states, there is a single edge mode at a sharp sample
boundary,\cite{Wen} so one can have a single pair of oppositely moving
modes running continuously through the transition region. We shall
discuss the transition between the two states in a narrow wire using a
fermion-Chern-Simons mean-field description\cite{HLR,Jain}, in which
the effective magnetic field changes sign in the transition region,
and using a bosonized Luttinger liquid formalism, in which the
interaction coefficient $g$ is allowed to vary continuously within the
transition region.  We also show that the existence of an adiabatic
junction between the two quantized Hall regions would allow
contruction of a DC step-up transformer, where the output voltage is
larger than the input voltage supplied by the power source.

\begin{figure}
\center
\epsfxsize=2.5truein
\hskip 0.0truein \epsffile{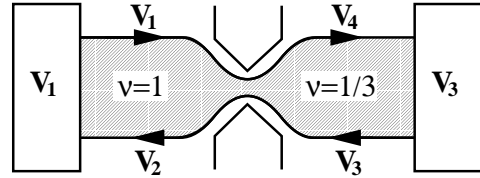}
\caption{Junction connecting quantum Hall states with
different filling factors $\nu=1$ and $\nu=1/3$. Quantum point contact
is produced by a narrow neck with the width of the order of the
 magnetic length. Arrows show the direction of the edge states.}
\label{fig:single}
\end{figure}

Consider the geometry illustrated in Fig.\ref{fig:single}, where there
is a narrow wire (or "point contact") connecting two macroscopic
quantized Hall regions, with different electron densities
corresponding to $\nu = 1$ and $\nu = 1/3$ respectively.  We assume
that each of the edges is sufficiently long that local thermal
equilibrium is established on the edge at a voltage labelled $V_j$,
where $j=1$ and $j=2$ denote, respectively, the incoming and outgoing
channels on the $\nu = 1$ side of the junction, and $j=3$ and $j=4$
denote the incoming and outgoing channels on the $\nu = 1/3$ side.
We also assume that the external current contacts are ``ideal'' so
$V_1$ and $V_3$ are equal to the voltages in the leads.\cite{BvH}

If the voltages of the external leads are equal to each other, then
 the system will be in global thermal equilibrium, with all $V_j$
being equal. More generally if $V_1-V_3$ is sufficiently small, the 
voltages $V_2$ and $V_4$ will be linear functions of $V_1$ and $V_3$,
 and we may write
\begin{eqnarray}
\label{alpha}
V_2 & = & \alpha V_1 + (1 - \alpha) V_3 ,\\
\label{beta}
V_4 & = & \beta V_1 + (1 - \beta) V_3 ,
\end{eqnarray}
where $\alpha$ and $\beta$ depend on the characterisitics of the
connecting junction, including, in general, the temperature $T$.

The current on edge $j$ is given by $I_j = \nu_j V_j (e^2/h)$, and the
energy flux along the edge is $ I_j V_j/2$.  Thus current conservation
through the junction requires that
\be
\beta = 3 (1 - \alpha) ,
\ee
while the requirement that the outgoing power be equal to or less than
the power incident on the junction implies
\be
1/2 \le \alpha \le 1
\ee

The two limiting situations, where there is no energy loss in the
junction region, are $\alpha = 1 , \beta = 0$, which corresponds to
zero current transmission through the junction, and $\alpha = 1/2 ,
\beta = 3/2$ , which is what we mean by an "adiabatic junction."  The
more familiar case of a wide junction, where equilibration is
established along a relatively long boundary separating bulk regions
with $\nu = 1$ and $\nu = 1/3$, coresponds to parameters $\alpha = 2/3
, \beta = 1$, which is not dissipationless.

If we set $V_3 = 0$, and supply a small voltage $V_1$ to the other
current lead, then a voltmeter connected between the opposite edges of
the $\nu=1/3$ wire will measure the voltage $V_4 = \beta V_1$.
Moreover, the two-terminal conductance $G$, defined as the ratio
between the current $I$ in the leads and the input voltage $V_1$, is
given by 
\be
G = \beta e^2 / 3h.
\ee
If we can construct a junction with $\beta > 1$, then we can obtain a
voltage $V_4$ which is larger than the input voltage, and we obtain $G
> e^2/3h$.  This last result violates the common belief that the two
terminal conductance of a quantum Hall system is always less than the
bottleneck with lowest conductance, as the two-contact resistance of
ideal leads connected to a single $\nu = 1/3$ region would be
$e^2/3h$. This also emphasizes an important point made by several
authors that the question of conductance is subtle and should be
formulated with a definite experimental arrangement in
mind.\cite{Matveev,Maslov,Alekseev}

\begin{figure}
\center
\epsfxsize=2.5truein
\hskip 0.0truein \epsffile{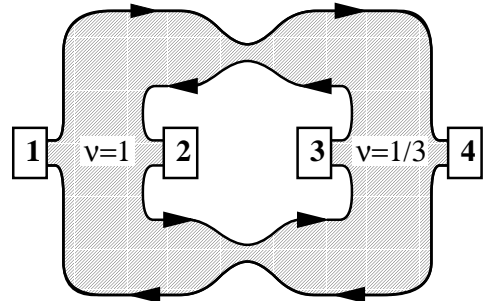}
\caption{Realization of the DC step-up transformer in the ring
geometry. Two quantum point contacts separate regions with different
filling factors. If a battery with voltage $V$ is attached to contacts
1 and 2 then the voltage drop between 3 and 4 can be $3V$ in the limit
of infinite load resistance.}
\label{fig:double}
\end{figure}

A more efficient voltage-transformer may be realized with the ring
geometry illustrated in Fig.\ref{fig:double}.  If a battery with
voltage $V$ is connected to ideal current contacts at points 1 and 2,
and a load with resistance $R$ is connected to points 3 and 4, then if
the junctions between the regions of $\nu = 1$ and $\nu = 1/3$ are
perfectly adiabatic ($\beta = 3/2$), the voltage across the load
resistor will be equal to $3V/(1+12h/e^2R)$.  When $R = \infty$, this
device draws no current from the battery, and the output voltage is
$3V$.  More generally, the output current is one-third of the input
current.  If $R \gg 12h/e^2$, the output voltage is close to $3V$, and
the power lost in the transformer is small compared to the power
delivered to the load.

To demonstrate the possibility of an adiabatic junction between $\nu =
1/3$ and $\nu = 1$ states, we first use the fermion-Chern-Simons
approach.\cite{HLR,Jain} In the mean field approximation the $\nu=1/3$
state is viewed as a completely filled Landau level for composite
fermions. This also holds for the $\nu=1$ state except that the
effective magnetic field is opposite to the direction of the external
magnetic field. Therefore, a narrow wire at either filling factor with
sufficiently sharp boundaries can be described in the Landau gauge by
a single energy band with two chiral edge channels. The two filling
factors can be easily distinguished in a wire much wider than the
magnetic length. In particular, the local electron density is three
times greater in the $\nu=1$ state. However, when the width of the
wire is of the order of the magnetic length the distinction between
the two states disappears. Then the density is not a good way to
differentiate between the states. In fact, on the mean field level the
two states look almost identical.

The transition between the two states can be carried out in the
following way. On one side we have a wide $\nu=1$ state with a single
energy band in the Landau gauge. The wire is then narrowed
gradually on the scale of the magnetic length. When the width of the
wire is of the order of the magnetic length the energy spectrum is
mainly determined by the confinement potential rather than magnetic
field. Therefore reducing the effective magnetic field along the
wire by reducing the density should not change radically the energy
spectrum. Higher composite fermion energy bands corresponding to other
fractions remain unfilled so that there is a single pair of edge
channels. As the filling factor is reduced below $1/2$ the effective
magnetic field changes sign and is slowly brought to its $\nu=1/3$
value. Then the wire is widened and represents a well-defined
$\nu=1/3$ state.

Although the composite fermion analysis can be extended to find the
chemical potentials of edge channels\cite{CH}, we take a different
approach here. It has been argued by several
authors\cite{Stone,Alekseev} that a quantum wire with filling factor
$\nu=1$ or $\nu=1/3$ can be modelled by a Luttinger Hamiltonian of
the form:

\begin{eqnarray}
\label{Lut}
H=\frac{\hbar}{4\pi}\int_{-\infty}^{+\infty}
\hspace{-0.5cm}v_F dx\left[\left(\frac{d\phi_L}{dx} \right)^2
+ \left(\frac{d\phi_R}{dx}\right)^2+\right. \nonumber\\
 \left. + \frac{g}{2}\left(\frac{d\phi_L}{dx}+
\frac{d\phi_R}{dx}\right)^2\right].
\end{eqnarray}
 We define charge-density operators $\rho_j$
 by $d\phi_j/dx=2\pi\rho_j$, and we assume commutation relations
\be
\label{com1}
[\phi_j(x),\phi_{j'}(x')]=(-1)^j i\pi{\rm sgn}(x-x') \delta_{jj'},
\ee 
where $j = 1,2$ refers to the indices $R$ and $L$, respectively.

In the $\nu=1$ state the density operators $\rho_j$ correspond to the
real electron density at a given edge, and $g=0$ for a sufficiently
wide wire.  In the $\nu=1/3$ state, however, the physical density at a
given edge is a linear combination of $\rho_j$, and $g=8$. By choosing
fields $\tilde{\phi}_j$ which are corresponding linear combinations of
$\phi_j$ one can get rid of the cross--term in the Hamiltonian and
express it in terms of decoupled edge excitations, as we shall see
explicitly below. Therefore, the presence of nonzero $g$ does not
necessarily imply interaction between the two edges but rather the
effective mixing of modes. The general relation between
$g$ and the filling factor valid for the simplest fractions, with a
single edge state, is:
\be
\label{gnu}
\nu=(1+g)^{-1/2},
\ee
where $\nu^{-1}$ must be an odd integer.\cite{Wen}

We take this idea further and postulate that the transition region
between the two regions with $g_1$ and $g_2$ can also be described by
the Luttinger Hamiltonian(\ref{Lut}) with $g=g_1$ for $x<-L/2$,
$g=g_2$ for $x>L/2$ and $g$ varying continuously from $g_1$ to $g_2$ for
$-L/2<x<L/2$.  This is a natural consequence of the fact that if the
translational invariance is not spontaneously broken in the wire and
the conditions in the wire are changed adiabatically then there are
two chiral boson modes running in opposite directions.

By using commutation relations $(\ref{com1})$ with the Hamiltonian we
 get the following equations of motion:
\begin{eqnarray}
\label{eqom}
\left\{ \begin{array}{c} \frac{d\phi_L}{dt} =
-v_F\left[\left(1+\frac{g}{2}\right)
\frac{d\phi_L}{dx}+\frac{g}{2}\frac{d\phi_R}{dx}\right]\\
\frac{d\phi_R}{dt} = v_F\left[\left(1+\frac{g}{2}\right)
\frac{d\phi_R}{dx}+\frac{g}{2}\frac{d\phi_L}{dx}\right]
\end{array}\right.
\end{eqnarray}
where $g$ and $v_F$ are functions of $x$.  The solution of these
equations depends on the particular form of $g$. However, there are
two limits when they can be solved exactly, independent of the way $g$
varies in the transition region.\cite{Oreg1} The first limit is when
the wavelength $\lambda$ of the incoming pulse is smaller that the
length $L$ of the transition region. In this case the solution can be
found by making a Bogoliubov transformation to chiral modes, which
correspond to the density waves confined to a single edge:
\begin{eqnarray}
\label{bogo}
\left\{\begin{array}{cc}
\tilde{\phi}_L=\frac{1}{2}(1+1/\sqrt{1+g})\phi_L+
\frac{1}{2}(1-1/\sqrt{1+g})\phi_R\\
\tilde{\phi}_R=\frac{1}{2}(1-1/\sqrt{1+g})\phi_L+
\frac{1}{2}(1-1/\sqrt{1+g})\phi_R
\end{array}\right.
\end{eqnarray}

 Substituting this in the equations of motion and neglecting the
variation in $g$ on the length of the pulse we find: 
\begin{equation}
\frac{d\tilde{\phi}_j}{dt}=(-1)^{j}v_F\sqrt{1+g}
\frac{d\tilde{\phi}_j}{dx}\\
\end{equation}
Solutions of these equations are chiral waves, which implies that a
short density pulse goes through the transition region without any 
reflection.\cite{Oreg1} In this sense we call the transition adiabatic.

The other limit is when the wavelength $\lambda$ of the incoming pulse
is greater than the length $L$ of the transition region. Then we can
solve the problem separately in the two regions and then match the
solutions assuming that the transition region is infinitely sharp.
The chiral wave solutions are found for the transformed variables
$\tilde{\phi}_L$ and $\tilde{\phi}_R$ with the values of $g$ in
Eq.\ref{bogo} corresponding to the particular regions. The continuity
of the $\phi_j$ requires that these fields are equal on the two sides
of the transition.

We formulate a scattering problem by forming an incoming wave with a
current of unit amplitude from the filling factor $\nu_1$ side. Then the
current in the reflected wave is given by the reflection coefficent
$r$ and the transmitted wave by the transmission coefficient $t$. We
find the current reflection and transmission coefficients:
\begin{eqnarray}
\label{coef1}
t & = & 2\nu_2/(\nu_1+\nu_2)\\
\label{coef2}
r & = & (\nu_1-\nu_2)/(\nu_1+\nu_2),
\end{eqnarray}
where $\nu_1$ and $\nu_2$ are related to $g_1$ and $g_2$ according to
Eq.(\ref{gnu}). It is easy to see that these coefficients satisfy the
law of current conservation: $r+t=1$. It also satisfies the law of
energy conservation. In fact the coefficients can be obtained from
these two conditions.  For the particular values $\nu_1=1$,
$\nu_2=1/3$, we find that the reflection coefficient is 1/2.  If the
incoming wave originates from the filling factor $1/3$ side
($\nu_1=1/3$, $\nu_2=1$), the reflection coefficient is $-1/2$. Minus
implies that the reflected pulse has the opposite sign of
density. The transmission coefficient is $3/2$ in this case, which may
appear to be a very unusual result. However, this is similar to a wave
reflection in a classical string with an impedance
discontinuity,\cite{Crawford} the impedance being the inverse of the
filling factor.\cite{Oreg2}

Knowing the reflection coefficients for the currents also allows us to
find edge state chemical potentials on the two sides of the transition
for DC transport. Let us send an infinite wavelength pulse from the
$\nu=1$ side with a current such that the voltage on that edge is
$V_1$ and a pulse from the $\nu=1/3$ with voltage $V_3$. Then the
outgoing currents can be found from Eqs.(\ref{coef1},\ref{coef2}). The
voltages on the outgoing channels are seen to obey
Eqs.(\ref{alpha},\ref{beta}), with $\alpha=1/2, \beta=3/2$.

Next, we consider the effects of disorder.  An impurity, or an
irregularity in the confining potential, at point $x$ in the
narrow-neck region can give rise to backscattering, through a term in
the Hamiltonian of form \be H' = \gamma {\rm exp} [i\phi_L(x) +
i\phi_R(x)] +{\rm h.c.}  \ee The phase of the coefficient $\gamma$
will depend on the position $x$, and its magnitude will depend
sensitively on the width of the strip at that point.  The amplitude
will be very small if $x$ is in a wide region, as there will then be
little overlap between the wavefunctions for states on opposite edges
of the wire.

The resistance due to backscattering is proportional to $|\gamma|^2$,
if $|\gamma|$ is small.  According to the standard renormalization
group analysis, however, for a wire of constant width, if $g > 0$,
the value of $|\gamma|$ will increase with decreasing energy scale.
Specifically, for voltages sufficiently small so that one is in the
linear regime, the backscattering resistance of a wire should
vary as $T^{-y}$, with \cite{Wen,KF}
\be
\label{scale}
y = (1+g)^{1/2} - 1.  
\ee
 For the present situation, where $g$ varies
with $x$, if the temperature is sufficiently high that the thermal
length scale $\hbar v_F/k_BT$ is small compared to the size $L$ of the
transition region, Eq.(\ref{scale}) still holds, with $g$ evaluated at
the position of the impurity.  The value of $y$ obtained in this way
would be intermediate between the values $y=0$ and $y=2$ that
correspond to uniform quantum Hall strips with $\nu=1$ and $\nu=1/3$
respectively.  If the temperature is sufficiently low that the thermal
length is large compared to L, however, then we find, from a
normal-mode analysis,\cite{CH} that the exponent $y$ becomes equal to 1,
independent of the precise location $x$ of the scatterer.\cite{foot}

In any case, we find that the adiabatic fixed point, where $\beta=3/2$
and there is no backscattering, is unstable, according to a
Luttinger-liquid analysis, so that any non-zero value of $(3/2-\beta)$
will grow with decreasing temperature and voltage.  Thus, to observe
the effect of voltage amplification, one must fabricate a junction
with a value of $(3/2-\beta)$ as close as possible to zero, and then
make the measurement at a temperature which is not too low.

There are several difficulties standing in the way of the experimental
implementation of the DC transformer. First, the quantum point
contacts must be approximately a magnetic length wide yet
adiabatic. Second, the edges of the $\nu=1$ and $\nu=1/3$ states must
be sufficiently sharp to support only a single edge channel.

In order to make the junction as close as possible to adiabatic, one
would like to avoid any roughness in the confining potential, as well
as impurities, which could lead to backscattering.  One must also
worry, however, about the possibility of an abrupt change in the
electron density or its profile across the width of the wire that
could occur due to a spontaneously-formed domain wall, if the electron
system goes through a first-order phase transition in the neck region.

Although we do not find any symmetry change between the $\nu=1/3$ and
$\nu=1$ states in a narrow wire, one cannot rule out the possibility
of having several phases separated by first-order transitions.  In
fact, exact-diagonaliziation studies of systems with up to six
electrons in a narrow wire suggest that there might be several
distinct phases, separated by sharp transitions, between the
densities which correspond to $\nu=1$ and $\nu=1/3$.
\cite{Yoshioka,Tapash} (The calculated states have different density
profiles across the wire, corresponding roughly to phases with one,
two or three distinct rows of atoms.)

Even if there is a sharp transition in a long wire, however, it might
be possible to obtain a smooth
transition in a properly engineered point contact.  Moreover, it is
possible in principle to cancel the reflected amplitude from one
density-discontinuity with a wave
reflected by a second discontinuity or by an impurity placed at an
appropriate position, using destructive interference.  Such a
 complicated structure may be difficult
to achieve by design, but might occur naturally in some fraction of
samples due to random flucuations during fabrication.
 
We are grateful to K.A. Matveev and to Y. Oreg for helpful discussions and to
T. Chakraborty for making Ref.\cite{Tapash} available prior to
publication. This work was supported by NSF grant DMR
94-16910 and by the Harvard Society of Fellows (for D.C.). The
hospitality of the Aspen Center for Physics, for D.C., and the Indian
Institute of Science, for B.H., is also acknowledged.


\widetext

\begin{thebibliography}{10}
\bibitem{Oreg2} Our result for $t$ differs from the related Eq.(10)
of Ref.\cite{Oreg1} due to a difference in definitions and an apparent
typographical error.
\bibitem{foot} If the voltage $V$ is large compared to $k_BT/e$, then
the temperature $T$ should be replaced by $eV/k_B$ throughout the
previous discussion.
\end{thebibliography}
\end{document}